# Why we live in the Computational Universe

## Giorgio Fontana

*Department of Information and Communication Technology, University of Trento, 38050 Povo (TN), Italy, 39-0461 883906, fax: 39-0461 882093, giorgio.fontana@unitn.it*

**Abstract.** To better understand the deep significance of our best physical theories it could be interesting to compare our Universe with its models. It may happen that the differences between the model and reality can be made indistinguishable, to the point that it may seem acceptable to consider reality as a gigantic program, a "mother computation" running in a Universal Computer. The computational interpretation of reality is here adopted for introducing concepts that are common in computer science, they may offer a new insight. For instance, code and memory usage optimization techniques are common in computer science because they improve the performances at a reduced hardware cost. According to the concepts discussed in this paper, the possibility of recognizing the effects of optimization rules in a physical reality will allow us to discriminate if our reality is fundamental or the result of a large computation. Conversely, code and memory optimization has side effects, if it is present in our Universe it can produce many interesting phenomena, some seem readily recognizable, others only wait to be discovered.

## INTRODUCTION

We all agree that our understanding and description of reality is only a collection of mathematical models applied to the properties of objects that we think are put in space by nature itself, an approximation of something that we cannot really comprehend or fully understand. Our science, and applications to space technology, has evolved by improving our abilities to measure, test and analyze physical events and by developing their mathematical models. The modeling approach is used not only by science but also by every living organism, whose behavior is determined by learning, because learning is the activity of making models. Human activity is directed towards learning the rules of the Universe and using them to our advantage. We are certainly successful within the confines of our intellectual limitations to the point that we have developed a technology that can be used to create self consistent Universes. In our schematization, Universes are composed by "space", "objects", "rules and procedures". For example a one dimensional Universe may be easily created with a Turing machine (Turing, 1936), that is the current technology in computer gaming, meteorology, many particle simulations and many other applications. The effective implementation is not important, the objects that lives inside the machine will never detect any difference in "their" behavior by changing the implementation.

Many scientists agree that we will never be able to understand if our reality is fundamental or, on the contrary, it resides in a higher (or lower) level reality. Even the fundamental dimensionality is uncertain. It may happen that the real space dimensionality is two, as some black hole theories suggest (Susskind, 1995). Thus, how do we live in a four dimensional structure that may in reality have only two-dimensions?

Is Physics only a collection of rules and methods, like it has been clearly stated for quantum mechanics? Is this a fundamental limitation to our understanding? This might not be true. Recent developments in Euclidean Relativity explain that our space-time is a process among many others in a more fundamental four-space (Fontana, 2005). The four-space is absolute and has common features in our current reality with the memory space of the Turing machine. The are two relevant features of the Turing machine: it is universal, as it defines all computable problems with independence from implementation; it has a simple structure, as it is composed by memory, processors and data structures. The universality of calculus has suggested to many scientist that our physical Universe can really be a computer simulation running on an unobservable Universal Computer (Hayes, 1999); philosophers are also involved in the research (Bostrom, 2003). Researchers recognize that the Universal Computer has all the power required to perform the job, and initially no further specification is given except that it defines the rules and runs the code. Fundamentally, the observation that our Universe and our space-time started with the Big Bang implicitly confirm that there is a problem of what existed before and may even exist in parallel with our Universe. The computational interpretation of the problem certainly simplifies the research because computer science has formalisms for dealing with similar, but much smaller, problems.

The first challenge is how to distinguish a "second level" or, with the language of computer science, simulated reality from the first level one belonging to the "First Level Universe" (FLU). The "second level" reality is named "Computational Universe", because of its similarity to computational models. Some scientists believe that the proof that we do indeed live in the Computational Universe can be discovered in past changes of physical constants (Barrow, 2002), hoping that they have been changed to better suit the anthropic principle valid in an Universe possibly designed for life.

In this paper we adopt a different approach, based on the language of computer science. The approach is based on the definition of the First Level Universe. "The First Level Universe is a system that is not optimized for efficiency and has no error". The First Level Universe must have a very simple structure to satisfy these conditions, therefore the Computational Universe is a much more complex place, being the complexity as its motivation.

We can therefore recognize if we really live in the Computational Universe because of the effects, consequences and limitations of the unavoidable code and memory optimization. The key concept is that every smart programmer optimizes the code he/she is producing, therefore if our reality appears optimized then it is NOT fundamental, alternatively if our reality appears not optimized then it is fundamental or it is not fundamental and the programmer is NOT smart, less smart than the worst student. It is certainly useful to recall that in computer science, strong optimization of the code is not without consequences, it usually produces rare unexpected results at boundary conditions, in addition optimization/compression of the data produces non-linearity and loss of "unnecessary" information in most cases, while in others we have the transfer of unwanted information. We therefore need to look for this phenomenology.

## EVIDENCE OF COMPUTATIONAL OPTIMIZATION

In computer science a process is a collection of predefined instructions executed on a data set that resides in a memory device. By comparing physical reality with the computational one, a particle can be considered a data-set that resides within the four-space, akin to a trace or a wave in the Euclidean four-space (Fontana, 2005). When a given particle is involved in an interaction, a specific "process" is executed on the particle, which makes it "alive" for the specific interaction, then the process is detached because of the economy of the computation, therefore particles have a defined state only if they are under "close" observation. It happens when there is a common process between the observed particle and the observer: incidentally this rule can be recognized as an important philosophical implication of Quantum Mechanics. Processes are conveniently recycled to compatible data sets and may have local variables; when a particle is destroyed in the four-space and then another particle immediately appears after, for instance, a tunneling process, it may inherit the same process according to the rules required to reside within the four-space.

If two particles have a "coupling process" that is efficiently managing a common property, then the process may change the property to both particles independent of the distance of the particles in the four-space (this is instantaneous action in space+time). This process should be impossible in classical physics, it is possible in quantum physics through particle entanglement but hardly explained in relativistic terms: the spooky action at a distance (Einstein, Podolsky, Rosen, 1935), but it is a trivial property of the "Computational Universe". The idea that a process is applied to a particle only when required, then the apparently instantaneous action at a distance can be interpreted as evidence of optimization as a general property of the Computational Universe. Simple background processes might be applied constantly to all particles like those due to the laws of gravity and inertia. We may extend the concepts straight to the most complex objects like organized matter or living beings. Those beings behave in a more complex way than as a sum of their parts, therefore an additional specific process has to be applied to make them work properly. When the editing of a complex object comes to a conclusion in the four-space, the managing process of the highest level activity is set free and not destroyed because it is not convenient to erase the process if the same process is required again. To test for the hypothesis it is only necessary to properly ask a living being if he/she has memory or recalls data or speaks languages from an alleged past life that are extraneous to the present history. It is not the scope of this article to find such a specific proof, but if common science ignores the result of many experiments because it cannot even define them, then we can now simply recognize that this is additional evidence of the optimization employed in the Computational Universe, and it is exactly the same kind of optimization that programmers and computer scientists usually apply to improve the performances of software: do not erase anything that is required (or could be required) again.

Quantum Mechanics appears to be a good description of the processes that are active on a given object, therefore it is the theory of the "process space". When a set of objects posses more properties than the summation of the parts, a new process has been attached to the set, the new process can have a quantum description: a state or a wave-function.

Till now we have talked about processes applied to data sets in the four-space. In Fontana (2005) we have inferred from Special and General Relativity that a space-time is comparable to an editing process in the four-space in which a form of past, present and future coexist, thus, the Euclidean view of Relativity takes the role of the theory of the "data space" of the Computational Universe. Many parallel space-times exist shifted along the proper time coordinate, preceding and following our present time (Fontana, 2005). Common experimental facts, if properly interpreted and understood, support the concept of the Computational Universe; "fields", that are basic building blocks of many theories, are completely immaterial, i.e. they are properties of the vacuum, and simply should not exist: only collisions with particles inside that space should be possible in plain space. In our reality, fields do indeed exist and this is a proof that mathematical models are the reality and the reality is the Computational Universe. A good programmer, dealing with a large four-dimensional data space which is almost empty, should use sparse matrix techniques to minimize the use of memory and minimum amount of data. Alternatively, variable and optimized four-space coordinate stepping should be used. With a fixed amount of memory available for hosting the four-space and using coordinate optimization techniques, it is not the size of the Universe that is limited, instead the number of particles (or traces) in the Universe should be limited, while the size of the Universe can be almost unrestricted, which is in agreement with observation. Empty space is but an illusion, it is not possible to occupy an arbitrary amount of space with objects because this will violate the memory conditions. Even distances are an illusion if space stepping is found to be non constant, this includes four-space, thus proper time stepping.

The most compelling evidence for the Computational Universe is the fact that physical laws and physical constants are the same everywhere in the Universe, how can this be possible even after the cosmological inflation if nothing has been discovered that keeps this information so precisely accurate across the entire Universe? (A particle or field which carries "physical laws" does not exist, nor can be experimentally removed.) Physical laws and constants simply are not a part of the four-space, they are written in the processes that compute the interactions among particles/objects.

Summarizing it is possible to identify the following main concepts:

1) The four-space of Euclidean Relativity is the inner view of the Universal Memory Space (UMS).
2) The UMS might be encoded, distances might be apparent.
3) Matter in our Space-time is the "data activity" of processes that edit the UMS. Many other Space-times may exist.
4) UMS is assumed to be a data space hosting pure information.
5) There are Processors and Processes (P&P), which are continuously reused, they are the hardware.
6) The real shape and nature of the combination UMS and P&P will probably become the new limit of our understanding.

## THE QUANTUM COMPUTATIONAL UNIVERSE

The critical view of physical laws enforce the concepts of the Computational Universe, which is not only a hypothesis for large scale phenomena but also a well known quantum gravitational theory. In fact a theory of the Quantum Computational Universe has been recently developed (the word Quantum has been added to distinguish from the classical one of the first part of the article: there is only one Computational Universe, the one in which we live). According to MIT scientist Seth Lloyd (Lloyd, 2005) the (Quantum) Computational Universe is a theory in which fundamental processes are described in terms of quantum information processing, with an explicit mechanism providing for the back-reaction or feedback of the metric to 'computational matter', black-hole evaporation, holography, and quantum cosmology. In the Computational Universe, 'quantum computation' is fundamental and all physics is derived from it. Similarly to causal sets (Bombelli, Lee, Meyer, Sorkin, 1987), there are causal structures that are directly related to the concept of space-time. There are also 'qubits', that are quanta of information and additional degrees of freedom used to derive the behavior of physical matter in a space-time.

The quantum information theory of the Computational Universe is substantially a consistent derivation of all observed phenomena from 'computation' going on in a Universal Quantum Computer (UQC) carrying the "mother computation". Quantum Mechanics is directly related to the inner mechanisms of the UQC, while GR can also be included.

The structure of space-time is derived from the behavior of qubits moving through the computation. Scattering between qubits correspond to an event and 'no scattering' is unobservable.

The superposition of scattering/no-scattering in the computation is a superposition of different causal structures, related to space-time metric fluctuations. In our Euclidean four-space view this corresponds to a possibility for a particle that is following a trace, to take a different route: the Computational Universe can be easily visualized as the UQC operating in the data space of Euclidean Relativity.

With such a powerful engine available, the discretized version of General relativity, Regge Calculus (Regge, 1961), has been adopted, which has been found to be compatible with the approach, in fact vertices of the embedded graph correspond to events and wires correspond to paths information takes in space-time. Because of the nature of quantum computation, any derived dynamical laws are generally covariant, thus implying Einstein's equations.

In the Einstein-Regge equation of the Computational Universe, computational matter tells space how to curve and space-time curvature tells qubits where to go, therefore the history of a time-line is determined by computation. As in the introductory part of this article, the hardware of the Universe is computational matter that 'we cannot see', what we see are the qubits, thus giving to our physical reality a quality of immaterial data space, exactly the opposite of what is commonly believed. The computational matter, hardware of the mother computer, might simply be "at rest" in the Euclidean four-space, enveloping the traces of all objects that evolve in the four-space. It is certainly possible to discover the presence of the computational matter and prove its possible existence within us by changing observer's speed in the four-space from $c$ to a lower speed due to changing the refractive index of the four-space (Fontana, 2005).

The reader might find common features in the computational matter rationale and in the "dark matter" of modern cosmology, this is only a possibility to be tested, computational matter could also be completely separate from data structures, like we have in our personal computers, and the Computational Universe might flawlessly work as described.

Assuming that the Computational Universe is made as an image of the First Level Universe, then the Computational Universe appears as an extended body in the space-time of the FLU. Taking into account that in a computer simulation, simulation quantities are unrelated to the "mother computer", hardware related, quantities that describe its state of motion in the FLU, the external quantities take their values in external dimensions. The external state of motion therefore adds seven additional parameters, three translations, three angles plus an additional absolute time. The seven parameters add six hidden space-like dimensions to the dimensionality of each particle of the Computational Universe and the additional absolute time parametrizes the evolution of the entire four-space. Summarizing, in the Computational Universe there are four large dimensions, six hidden dimensions and an absolute time.

## OBSERVABLE FEATURES OF THE COMPUTATIONAL UNIVERSE

According to reference article (Lloyd, 2005), of which we report here a few results, it should be possible to recognize the validity of the Computational Universe by looking for its consequences that are:

1) Quantum fluctuation in the routing of the signal through the computation can produce gravitational waves. Situations in which strong quantum computation is performed should produce a significant amount of gravitational waves
2) Initial singularities correspond to adding bits to the computation.
3) Interaction of new bits with another bit crates a new volume of space-time.
4) Final singularities correspond to when bits leave the computation.
5) It is possible to have temporary black holes in absence of singularities, when a 'process' of the computation is temporarily sequestered from the computation. When it rejoins the rest of the computation, there appears an apparent horizon that later disappears. The geometric quantum limit (Giovannetti, Lloyd, Maccone, 2004) limits the number of elementary events in a four-volume. This implies the holographic principle.
6) The number of elementary events and bits in the Universe appears to be less or equal to $10^{123}$. The average spacing between elementary events in the Universe is about $10^{-13}$ s, therefore there are places (like here on Earth) in which this spacing is much smaller, and others in which this spacing is much larger. The computational efforts seem to adapt to the difficulty of the specific situation, similarly to variable step simulation of most engineering software. It could be interesting to find proof of this behavior as result of an experiment, a positive result will show that our reality is not fundamental, but the product of a large computation. Alternatively, the local "intensity" of the computation might depend on the local complexity of the problem (how many particles are participating to an event), therefore it determines $c$ in the Euclidean four-space, thus it determines the refractive index of the four-space. The refractive index of the four-space can be formally related to curvature in a space-time (Fontana, 2005) at least in a number of typical situations. In the Computational Universe, "computation" is the source of the gravitational field. It is reasonable that the complexity of a simulation problem is proportional to the relativistic total energy of the system being simulated, on the other hand it is similarly reasonable that the FLU may create gravity non only by managing the basic evolution of particle motion but also by managing higher level tasks like those discussed in the second section of this article.
7) Any quantum computer can simulate another quantum computer, resulting in possible nesting of Universes of various level.

The memory and computational limitations are similar to those derived from the classical view of the Computational Universe.

Recent experiments add to the concept that our reality develops through computations and that exceedingly difficult problems make the UQC to give up the task and to create a new wavefunction to make the problem computable. If two lasers in cascade are made to interact, then their fluctuations are transferred from the first to the second after some delay. But using three lasers, with propagation delay, the outer two are observed to be in perfect synchronization. If coupled lasers can be described with a finite number of equations, coupled lasers with delay require an infinite number of equations: the UQC gives up the task and "creates" a new wavefunction, which is obviously instantaneous, leading to perfect synchronization (Cho, 2006). We therefore conclude that, if required, new wavefunctions are created to keep the complexity low. The laser test may be used to probe to computational abilities of the UQC and to identify its strategies.

## CONCLUSION

It could be actually impossible to prove with absolute certainty if our reality is fundamental or the result of a large computation, but numerous observations indicate that the same tricks and limitations we usually find in computer science applications are also present and sometimes plainly evident in physical laws and in still unresolved observations. Further research is certainly required, nonetheless a scientific methodology has been developed that should enable us to discriminate if we live in a fundamental reality or in the "Computational Universe".

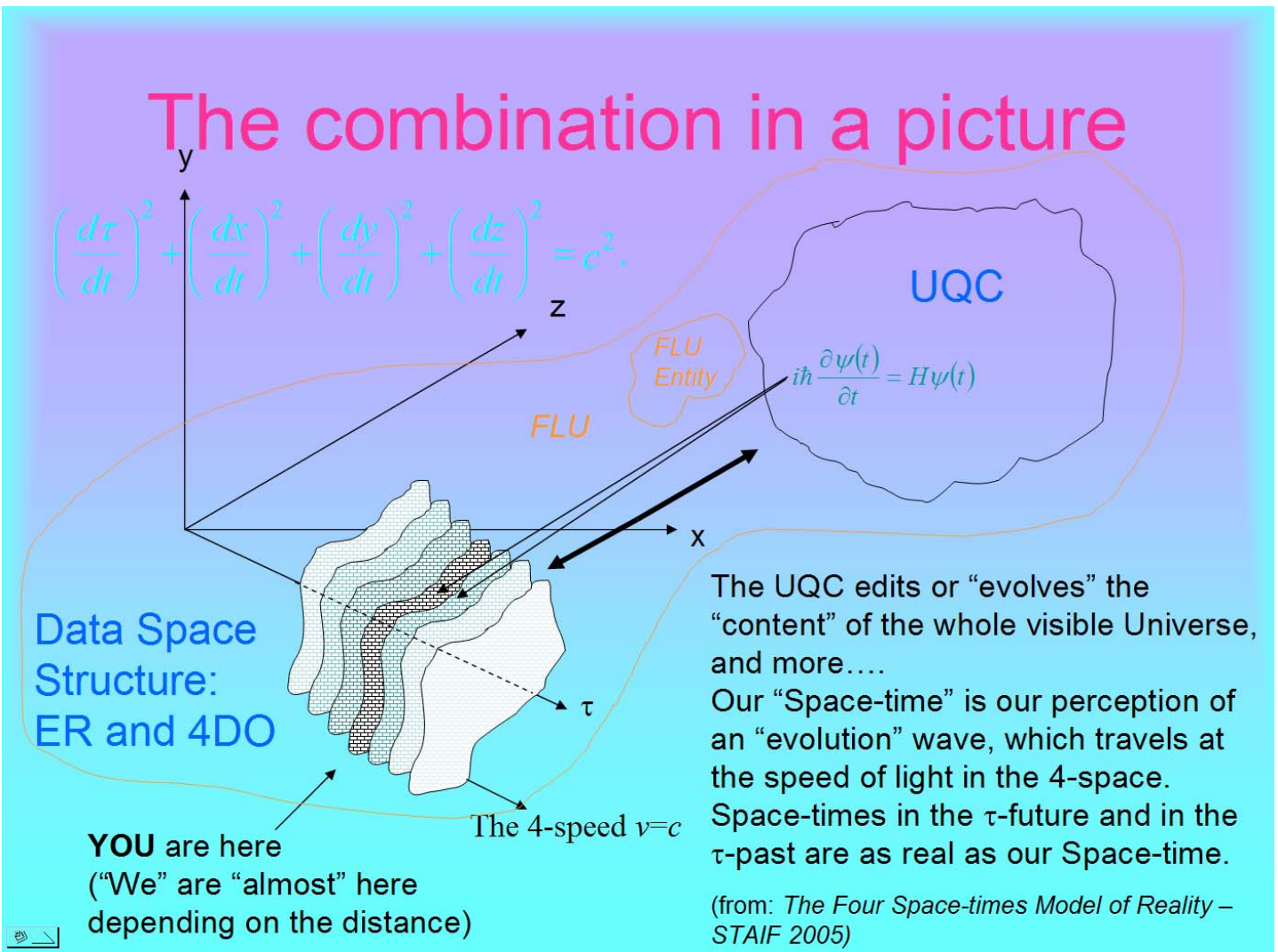